\documentstyle[emulateapj]{article}

\catcode`\@=11 % This allows us to modify PLAIN macros.
\def\@versim#1#2{\vcenter{\offinterlineskip
        \ialign{$\m@th#1\hfil##\hfil$\crcr#2\crcr\sim\crcr } }}

\newcommand{\beq}{\begin{equation}}
\newcommand{\eeq}{\end{equation}}
\def\lsim{\mathrel{\mathpalette\@versim<}}
\def\gsim{\mathrel{\mathpalette\@versim>}}

\def\md{\dot m}

\def\ri{r_{\rm in}}

\begin{document}

\title{Possible Evidence for Truncated Thin Disks in the
  Low-Luminosity Active Galactic Nuclei M81 and NGC 4579}
\author{Eliot Quataert\footnote{Chandra Fellow; present address:  Institute for Advanced Study, School of Natural Sciences, Olden Lane, Princeton, NJ 08540; eliot@ias.edu}, 
Tiziana Di Matteo\footnote{Chandra Fellow},
  Ramesh Narayan} 
\affil{Harvard-Smithsonian Center for Astrophysics,
  60 Garden St., Cambridge, MA 02138; equataert, tdimatteo, rnarayan,
  @cfa.harvard.edu} 
\author{and} 
\author{Luis C. Ho} 
\affil{Carnegie Observatories, 813 Santa Barbara St. Pasadena, CA 91101-1292; lho@ociw.edu} 
\medskip

\begin{abstract}
  M81 and NGC 4579 are two of the few low-luminosity active galactic
  nuclei which have an estimated mass for the central black hole,
  detected hard X-ray emission, and detected optical/UV emission.  In
  contrast to the canonical ``big blue bump,'' both have optical/UV
  spectra which {\em decrease} with increasing frequency in a $\nu
  L_\nu$ plot.  Barring significant reddening by dust and/or large
  errors in the black hole mass estimates, the optical/UV spectra of
  these systems require that the inner edge of a geometrically thin,
  optically thick, accretion disk lies at $\sim 100$ Schwarzschild
  radii.  The observed X-ray radiation can be explained by an
  optically thin, two temperature, advection-dominated accretion flow
  at smaller radii.  \ 

\

\noindent {\em Subject Headings:} accretion, accretion disks -- 
black holes -- galaxies:  individual (M81, NGC 4579)

\end{abstract}
 
\section{Introduction}
%A recent spectroscopic survey shows that nearly half of all nearby
%galactic nuclei have emission line properties suggestive of
%low-luminosity active galactic nuclei (LLAGN; Ho, Filippenko, \&
%Sargent 1997).  This identification is supported by growing dynamical
%evidence that the majority of nearby early-type galaxies contain
%central ``massive dark objects,'' most likely supermassive black holes
%(Kormendy \& Richstone 1995; Magorrian et al.  1998; Ho 1998).
%Accretion onto a central supermassive black hole may therefore be
%commonplace in galactic nuclei.

There are several theoretical models for the structure of matter
accreting onto a central black hole, e.g., spherical accretion (Bondi
1952), geometrically thin disks (Shakura \& Sunyaev 1973; see Frank,
King, \& Raine 1992 for a review) and their associated coronae (e.g.,
Haardt \& Maraschi 1991), and advection-dominated accretion flows
(ADAFs; Ichimaru 1977; Rees et al.  1982; Narayan \& Yi 1994;
Abramowicz et al. 1995; see Narayan, Mahadevan, \& Quataert 1998, and
Kato, Fukue, \& Mineshige 1998 for reviews).  One of the fundamental
questions in accretion theory is to what degree these models are
relevant for understanding observations of black holes in galactic
nuclei and X-ray binaries.

Lasota et al. (1996) suggested that the supermassive black hole in NGC
4258 and most low-luminosity active galactic nuclei (LLAGN) accrete
via an ADAF, with any thin disk restricted to large radii.  This is a
natural identification since ADAFs can only exist in sub-Eddington
systems ($L \lsim 0.01-0.1 \ L_{\rm edd}$) and their low radiative
efficiency may help explain the observed low luminosities.

Part of the difficulty in observationally testing Lasota et al.'s
hypothesis is that many LLAGN have hard X-ray continuum properties
very similar to those of luminous Seyferts, namely photon indices
$\approx 1.8$ (Serlemitsos, Ptak, \& Yaqoob 1996).  Hard X-ray
continuum observations are therefore not necessarily the best
discriminant among theoretical models; what is required to reproduce
the observations is an appropriate Comptonizing medium, which both
ADAFs and accretion disk coronae can provide.

One must therefore look to different wavebands to more definitively
test theoretical models.  For example, infrared and optical/UV
observations provide excellent probes of the blackbody emission
expected from optically thick, geometrically thin, accretion disks
(the ``big blue bump''; see, e.g., Koratkar \& Blaes 1999 for a
review).

%For NGC 4258,
%even within the context of an ADAF model, a thin accretion disk
%outside of $\sim 30$ Schwarzschild radii is required by infrared
%observations (Gammie, Narayan, \& Blandford 1999).

%Radio observations of proposed ADAF systems are particularly useful as
%self-absorbed synchrotron emission in the radio is a robust prediction
%of the model (Narayan \& Yi 1995b).  Di Matteo et al. (1999ab,
%hereafter DM99ab; see also Quataert \& Narayan 1999; QN) carried out
%such a test in a number of cluster ellipticals (e.g., NGC 4649).
%Their radio data are incompatible with the canonical ADAF model of
%Narayan \& Yi (1995b); specifically, the models predict too much flux.
%In addition, the observed X-ray to radio flux ratio and X-ray spectral
%index (Allen et al. 1999; DM99ab) both point towards additional
%physics in the accretion flow, perhaps that most of the mass supplied
%to the accretion flow is lost to an outflow/wind rather than reaching
%the central black hole (Blandford \& Begelman 1999).

In this Letter we investigate the mode of accretion in LLAGN by
comparing theoretical models with the broad band spectra of the nuclei
of M81 and NGC 4579.  These systems are rough analogues of the well
studied maser galaxy NGC 4258.

%In the next section we briefly review the observations (\S2).  In \S3
%we discuss our theoretical models and in \S4 we compare them with the
%data.  In \S5 we conclude and place our results in the larger context
%of LLAGN theory.

\section{The Observations}

The criteria we have used in choosing systems to investigate are the
following: (1) There is a mass estimate for the central supermassive
black hole; in M81 this is a dynamical estimate from stellar
kinematics (Bower et al.  1996) while in NGC 4579 the estimate is
based on modeling the BLR (Barth et al.  1996).\footnote{Note that
this technique worked quite well for M81 (Ho, Filippenko, \& Sargent
1996).}  (2) There are high resolution spectral observations of the
nucleus with VLBI, {\it HST}, ground based IR, and {\it ASCA},
supplemented by {\it ROSAT}; references to the original data are given
in the compilation by Ho (1999; hereafter H99).

%M81 and NGC 4579 are not the only systems which satisfy these criteria
%(e.g., M87 does; see, e.g., Di Matteo et al. 1999b) but we believe
%that, together with NGC 4258, they form an important and distinct
%subset of LLAGN (see \S5).

%  new constraints on the theoretical models come from
%the IR and optical/UV data (HST), which constrain, e.g., the
%properties of geometrically thin accretion disks.\footnote{The
%  constraints on M87 imposed by Allen et al's (1999) observations are
%  crucial, but not ``new'' in the present context because they are
%  discussed in detail by DM99b.}  The optical/UV data are subject to
%two significant uncertainties: stellar contamination and dust.  The
%importance of stellar contamination has been estimated from the
%strength of absorption lines in the optical/UV spectrum.  Dust is not,
%however, so readily accounted for.  Throughout the bulk of this paper
%we interpret the data assuming negligible extinction; the possible
%importance of dust is discussed in the Appendix (see also \S4 of H99).

\section{Theoretical Models}

%We consider two general classes of theoretical models for the observed
%systems, geometrically thin, optically thick accretion disks (plus
%coronae) and ADAFs (plus outer thin disks).

We measure black hole masses in solar units and accretion rates in
Eddington units: $M = m M_{\odot}$ and $\dot M = \dot m \dot M_{edd}$.
We take $\dot M_{edd} = 10L_{edd}/c^2 = 2.2 \times 10^{-8} m M_{\odot}
{\rm yr}^{-1}$, i.e., with a canonical 10 \% efficiency.  We measure
radii in the flow in Schwarzschild units: $R = r R_s$, where $R_s =
2GM/c^2$.
 
We model emission from a thin accretion disk as a multicolor blackbody
(e.g. Frank et al. 1992). The emission is independent of the
microphysics of the disk and depends primarily on $m$, $\md$, and the
inner edge of the disk ($\ri$).\footnote{In constructing the disk
  models, a boundary condition must be applied at $\ri$; we choose the
  standard zero torque boundary condition for which the disk
  temperature goes to zero at $\ri$; whether this is correct for $\ri
  \ne 3$ is unclear.  It is, however, the conservative approach, as a
  number of our conclusions will involve limits on $\ri$.  Alternative
  boundary conditions would give larger limits on $\ri$.}  We take the
outer edge of the disk to be at $r_{\rm out} = 10^5$ and the
inclination of the disk to the line of sight to be $i = 60^o$, but the
choices are not crucial.  Multicolor blackbodies are known to be a
poor approximation to big blue bumps observed in, e.g., quasars (see
the recent review by Koratkar \& Blaes 1999 and references therein).
For our purposes, however, they suffice, as we seek only a crude
estimate of the presence/absence of a thin disk in particular systems.
Corrections due to irradiation of the disk by the X-ray source or
electron scattering in the disk atmosphere are insufficient to change
our general conclusions.

In models in which the thin disk is truncated at $\ri > 3$, we model
the accretion on the inside as an ADAF (for related models of black
hole X-ray binaries see, e.g., Esin, McClintock, \& Narayan 1997).
The outer radius of the ADAF is assumed to be equal to the inner
radius of the thin disk, $\ri$, corresponding to a thin disk
``evaporating'' (see, e.g., Meyer \& Meyer-Hofmeister 1994; Honma
1996) to form an ADAF.

In contrast to a thin disk, the predicted spectrum from an ADAF
depends on several microphysics parameters, notably the ratio of gas
to magnetic pressure, $\beta$, the viscosity parameter, $\alpha$, and
the fraction of the turbulent energy in the plasma which heats the
electrons, $\delta$.  In this paper, we fix $\alpha = 0.1$, $\beta =
10$, and $\delta = 0.01$ (for motivation, see Quataert \& Narayan
1999).  We adjust $\md$ to reproduce the observed X-ray flux.  

%As
%discussed below, we adjust $\ri$ to explain the available optical/UV
%data.

%We model the spectra of ADAFs following Quataert \& Narayan (1999) and
%references therein.\footnote{Note that the Quataert \& Narayan paper
%  focuses on the observational implications of substantial mass loss
%  from an ADAF.  For simplicity we do not consider mass loss in this
%  paper (we have confirmed that our conclusions are unchanged by its
%  inclusion).}  The predicted spectrum consists of thermal
%synchrotron, Compton, and bremsstrahlung emission and depends on $m$,
%$\md$, and several parameters which characterize the microphysics of
%the accretion flow:

% and the ``outflow parameter,'' $p$.  The
%parameter $p$ is a power law index for the accretion rate in the flow
%(i.e., $\md \propto r^p$) and accounts for the possibility that
%significant mass may be lost to an outflow/wind.  The radius $r_p$ is
%defined to be the radius inside of which mass loss takes place (i.e.,
%the power law for $\md$ only applies inside $r_p$).  This need not be
%equal to the outer radius of the ADAF, as winds could, in principle,
%only become important well inside the outer boundary.  For ADAF models
%with winds, we quote values of $\md$ at radii $\gsim r_p$.  This
%corresponds to the rate at which matter is being fed to the accretion
%flow.

\section{Comparison with the Data}
%We discuss theoretical models for M81, NGC 4579, M87, and NGC 4594 in
%the following subsections.  We focus primarily on M81 and M87, with
%briefer discussions of NGC 4579 and NGC 4594.

%\subsection{M81}

Figure 1 shows the available data for M81 ($m = 4 \times 10^6$)
compiled by H99. There are radio detections, IR limits, an optical/UV
detection with {\it HST} (the two points at $\approx 3 \times 10^{15}$
Hz are due to variability in the source), and an {\it ASCA} detection
in X-rays.

The dotted lines in Figure 1a show geometrically thin accretion disk
models with (from top to bottom) $\md = 10^{-2}, 10^{-3}, 3 \times
10^{-4}$, and $3 \times 10^{-5}$.  The disks are assumed to extend all
the way to the marginally stable orbit for a Schwarzschild black hole
($\ri = 3$).  None of these models is capable of explaining the {\it HST}
optical/UV data.  In particular, the model with roughly the correct
flux level ($\md \approx 3 \times 10^{-4}$) is far too hot while the
model which fits the higher frequency {\it HST} data ($\md \approx 3 \times
10^{-5}$) cannot account for the lower frequency {\it HST} data.

The solid line in Figure 1a is a thin disk model with $\md \approx
10^{-2}$ and $\ri \approx 100$, corresponding to a thin disk
``truncated'' at about 100 Schwarzschild radii.  This model reproduces
the lower frequency {\it HST} data fairly well.  It does somewhat
underproduce the higher frequency data, but this is not a serious
difficulty because the Compton emission necessary to account for the
X-rays can remove this discrepancy (see Fig. 1b).

Note that for fixed $m$, $r_{\rm out}$, and $i$, the thin disk model
depends on two parameters, $\md$ and $\ri$.  In turn, the observations
provide, roughly speaking, two numbers, the luminosity and the maximum
temperature of the disk.\footnote{The latter follows because the UV
  slope in M81 requires the {\it HST} data to be on the exponential tail of
  the disk model (which requires $k T_{\rm max} \sim h \nu_{min}$,
  where $\nu_{min} \approx 3 \times 10^{14}$ Hz is the minimum
  frequency of the {\it HST} observations).}  This allows one to fit a
unique model to the data and determine, to within a factor of few, the
values of $\md$ and $\ri$.

If the above interpretation of the M81 data is correct, it represents
good evidence for a truncated disk in a galactic nucleus. This would
have important implications for accretion theory.  It is therefore
worth emphasizing the assumptions made in our analysis.  First, we
assume that the black hole mass in M81 is relatively accurately
determined to be $m \approx 4 \times 10^6$ (see Bower et al. 1996 for
the dynamical estimate, which is somewhat uncertain).  If the true
mass were to be larger by a factor of $\gsim 10$, the thin disk
emission would be colder ($T \propto m^{-1/4}$) and would therefore be
capable of explaining the observations (we find, e.g., that $m = 4
\times 10^7$, $\md = 10^{-5}$, and $\ri = 3$ can account for the
observed optical/UV spectrum).

The interpretation of the optical/UV data is subject to two additional
uncertainties: stellar contamination and dust.  The importance of
stellar contamination has been estimated from the strength of
absorption lines in the optical/UV spectrum.  Dust is not, however, so
readily accounted for.  H99 discusses the possibility that reddening
by dust can account for the unusual optical/UV emission in M81, NGC
4579, and a number of other LLAGN.  He argues that such an explanation
is unlikely, but cannot be fully excluded.  In particular, any such
dust must have properties different from Galactic or SMC dust;
moreover, the presence and/or properties of such dust must be a strong
function of the luminosity of the AGN (since luminous AGN do not show
optical/UV spectra similar to M81).  Until better IR limits are
available to constrain reemission by dust, it is unlikely that this
question will be fully settled. Throughout this paper we interpret the
data assuming negligible reddening.

It is important to note that the interpretation of the optical/UV
emission from M81 is not sensitive to uncertainties in our modeling of
the thin disk emission.  More sophisticated disk models generally have
color temperatures which exceed the effective temperature (e.g.,
Shimura \& Takahara 1995), implying that the emission is {\it hotter}
than a blackbody, not {\it colder} as is needed to explain the
observations of M81.

Figure 1b shows a model of M81 in which the thin disk is truncated at
$\ri \approx 100$, inside of which there is an ADAF ($\md$ is $\approx
10^{-2}$ in both). The solid line shows the total emission from this
model while the dashed line isolates the ADAF contribution.  This
combined ADAF + disk model provides a good description of the
optical/UV and X-ray data. In particular, the X-ray slope is
reproduced fairly well by the ADAF model, without tuning any
additional parameters.  The X-ray emission from the ADAF is due to
Compton scattering of synchrotron photons produced by thermal
electrons in the near equipartition magnetic field of the accretion
flow.  For $\ri \sim 100$ and for typical temperature and density
profiles in ADAFs, the soft photons produced by the disk at $r \gsim
100$ are negligible compared to those produced by synchrotron emission
in the ADAF.

More generally, the X-ray data in M81 can be explained by any
appropriate ``Compton cloud'' inside the truncated disk (as in models
for the hard state of galactic black hole candidates; see, e.g.,
Poutanen, Krolik, \& Ryde 1997).  The difference between this and the
ADAF model is (1) the source of soft photons for Comptonization is
taken to be the disk emission, rather than synchrotron emission and
(2) there is no dynamics for the cloud, just a specification of its
Thomson optical depth $\tau$ and electron temperature $k T_e$.  The
X-rays can also be produced by a corona on top of a thin disk which
extends all the way to $\ri = 3$.  This highlights the issue raised in
the introduction: hard X-ray continuum observations cannot always
discriminate between ADAF and coronae models.  The optical/UV
observations discussed above, while they have their own uncertainties,
are necessary for further testing theoretical models.

The observed radio emission from M81 cannot be explained by the models
considered here.  It is probably synchrotron emission by nonthermal
electrons in an outflow/jet (there is good evidence for a jet in M81;
Bietenholz et al. 1996; Ho et al. 1999).

%\subsection{NGC 4579}

Figure 2 shows the available data for NGC 4579 ($m \approx 4 \times
10^6$).  The broad band spectrum is rather similar to that of M81
(although the data are not as good).  There is evidence for a declining
optical/UV spectrum (with variability) and a canonical ``Seyfert''
X-ray photon index of $\approx 1.7$.  As in M81, the optical/UV
spectrum suggests a truncated disk.  The dotted lines in Figure 2a are
for thin disk models with $\ri = 3$ and $\md = 3 \times 10^{-2}, \ 10^{-3}$, and $10^{-4}$, none of which can account for the optical/UV
data.  The solid line in Figure 2a is a disk model with $\md \approx
0.03$ and $\ri \approx 100$, which accounts qualitatively for the
observations.

Figure 2b shows an ``ADAF + disk'' model of NGC 4579 with $\md \approx
0.03$ and a transition radius of $\ri \approx 100$. As in M81, this
reproduces the data reasonably well, although the higher frequency UV
data are somewhat underproduced.

\section{Discussion}

Our principal result is that the optical/UV to X-ray emission detected
from the nuclei of M81 and NGC 4579 can be well explained by a model
in which an optically thick, geometrically thin accretion disk extends
down to $\approx 100$ Schwarzschild radii, inside of which lies an
optically thin, two-temperature, ADAF.  Such an interpretation is
%independent of the detailed physics of the ADAF or ``Compton cloud''
%inside $\approx 100$ Schwarzschild radii; it is also 
relatively independent of ``greybody'' effects which lead to
deviations of the disk emission from a multicolor blackbody.  It does,
however, assume that dust in the galactic nuclei of interest is
unimportant in modifying the intrinsic optical/UV spectrum of the
LLAGN and that current black hole mass estimates are accurate to order
of magnitude.  These assumptions should be investigated in more detail
in future work.

%Modulo the above assumptions, the inference that the thin disks in M81
%and NGC 4579 are truncated at $\approx 100$ Schwarzschild radii is
%robust.  It follows directly from the observed optical/UV spectra,
%which decrease with increasing frequency in a $\nu L_\nu$ plot.  

Our models can be tested by future X-ray spectroscopy with the {\it
CXO} and {\it XMM}.  Detection of an iron line with large relativistic
broadening in M81 or NGC 4579 would argue strongly against a truncated
thin accretion disk. {\it ASCA} observations in fact detect an iron
line in these systems, but with a centroid energy of $\approx 6.7$ keV
(see Ishisaki et al. 1996 for M81 and Terashima et al. 1998 for NGC
4579).  The M81 line is broadened by $\sim 10 \%$, while there is no
evidence for a broad line in NGC 4579. This high energy emission line
is incompatible with the $6.4$ keV fluorescent line expected from a
canonical thin accretion disk.  It is possible that such a line could
arise in the transition region between an ADAF and a disk, but it is
unclear whether one can simultaneously have the transition region
subtend sufficient solid angle with respect to the X-ray emitting
plasma and be partially, but not fully, ionized.  If so, the line
should be broadened by $\approx 10 \%$ (but not by more).

%\subsection{Discussion}

To conclude, we briefly place the present results in the context of
other recent work on LLAGN.
 % ADAFs can only exist below a critical
%accretion rate, $\dot m_c \approx \alpha^2$ (e.g., Rees et al. 1982).
%An important question in accretion theory is whether $\dot m$ alone
%determines the mode of accretion (see, e.g., Narayan \& Yi 1995 for a
%discussion of these issues).  In one view, if $\dot m \gsim \dot m_c$,
%accretion proceeds via a thin disk, while if $\dot m \lsim \dot m_c$,
%accretion proceeds via an ADAF in the inner parts of the flow, with a
%thin disk restricted to the outer parts of the flow.  The transition
%between the two occurs at $\ri$, which is taken to be a function of
%$\dot m$ only.  An alternate view (e.g., Begelman 1986) argues that
%the mode of accretion depends not only on $\dot m$, but also on the
%outer boundary conditions supplied to the flow (e.g., the specific
%angular momentum and temperature of the inflowing material).  We
%believe that the present paper, together with additional recent work
%on LLAGN, argues for the importance of boundary conditions.
Gammie, Narayan, \& Blandford (1999) argued that an ``ADAF + disk''
model gives a reasonable description of the broadband spectrum of NGC
4258; they showed that recent IR observations require that the
transition radius between the ADAF and the disk lies at $\approx
10-100$ Schwarzschild radii.  The $\dot m$ of their model is $\approx
0.005$ (for $\alpha = 0.1$).  There is thus reasonable consistency
between the models of M81 and NGC 4579 presented in this paper and
Gammie et al.'s model of NGC 4258: as expected theoretically, the
models have comparable values of $\ri$ for comparable $\dot m$.
%Note that for values of $\alpha \approx
%0.1-0.3$, $\dot m \approx \dot m_c$ in M81, NGC 4579, and NGC 4258.

By contrast, Di Matteo et al. (1999ab) constructed ADAF models to
explain the broad band spectrum of 6 elliptical galaxies in clusters.
They argue that values of $\dot m \approx 0.03$ at large radii are
needed to explain the spectra of, e.g., M87 and NGC 4696.  We have
compared thin accretion disk models at this $\dot m$ with the observed
spectra of M87 and NGC 4696.  We find that any thin disk must be
truncated at $\gsim 10^4$ Schwarzschild radii in order not to violate
limits on the IR emission from these systems.  The limits on $\ri$ in
the cluster ellipticals therefore exceed the values estimated in M81,
NGC 4579, and NGC 4258 by at least two orders of magnitude.

A plausible explanation for this difference in $\ri$ at comparable
$\dot m$ is the importance of boundary conditions in determining the
mode of accretion (cf Begelman 1986).  In the ellipticals, a cooling
flow supplies hot, roughly spherically symmetric gas to the accretion
flow.  The gas remains in this state as an ADAF all the way to the
black hole.  By contrast, in M81, NGC 4579, and NGC 4258, the
interstellar medium feeding the accretion flow may reside primarily in
a disk.  The accretion can proceed via an ADAF on small scales only
after the disk has ``evaporated'' into an ADAF (by a still poorly
understood process).  
%This difference in the formation process of the
%ADAF, dictated ultimately by the boundary conditions at large radii,
%manifests itself in different values of $\ri$.

%The hypothesis that the mode of accretion in LLAGN depends on both
%$\dot m$ and the boundary conditions of the flow can also be tested by
%future X-ray spectroscopy (not to mention by a larger sample of broad
%band spectra and black hole mass estimates).  It predicts that, in a
%sample of LLAGN, there should be no correlation between the equivalent
%and/or physical line width of the fluorescent iron line and the
%Eddington luminosity of the source.

\acknowledgements We thank Eric Agol, Anthony Aguirre, Aaron Barth, Julian 
Krolik, Aneta Siemiginowska, and Marco Spaans for useful discussions.
We acknowledge support from an NSF Graduate Research Fellowship (EQ),
NSF Grant AST 9820686 (RN), and NASA/STScI grants GO-06837.01-95A,
AR-07527.02-96A, and AR-0831.02-97A (LCH).  TDM (EQ) acknowledges
support provided by NASA through Chandra Fellowship grant number
PF8-10005 (PF9-10008) awarded by the Chandra X-ray Center, which is
operated by the Smithsonian Astrophysical Observatory for NASA under
contract NAS8-39073.

%\begin{multicols}{2}

\newpage

\begin{figure}
\plottwo{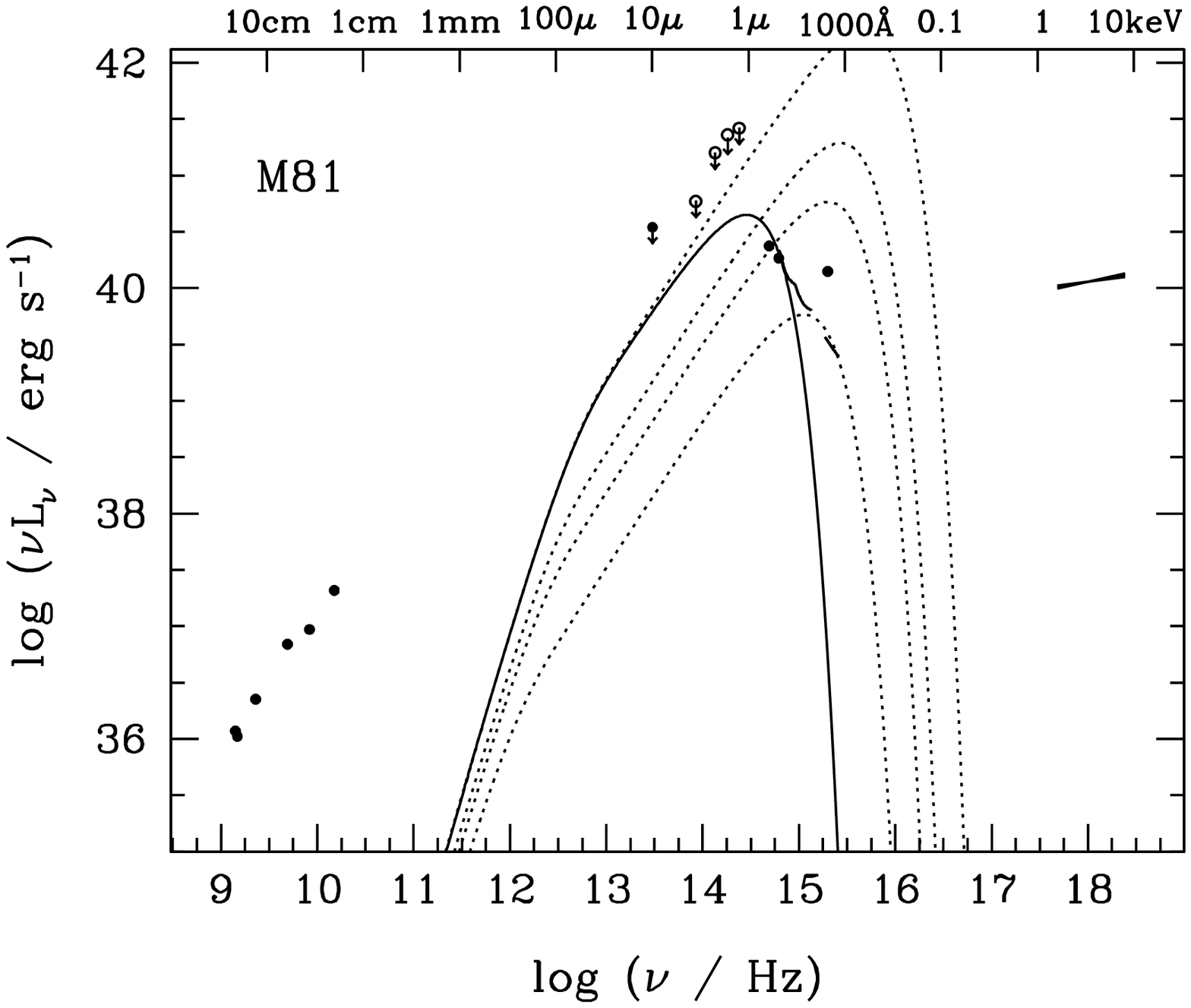}{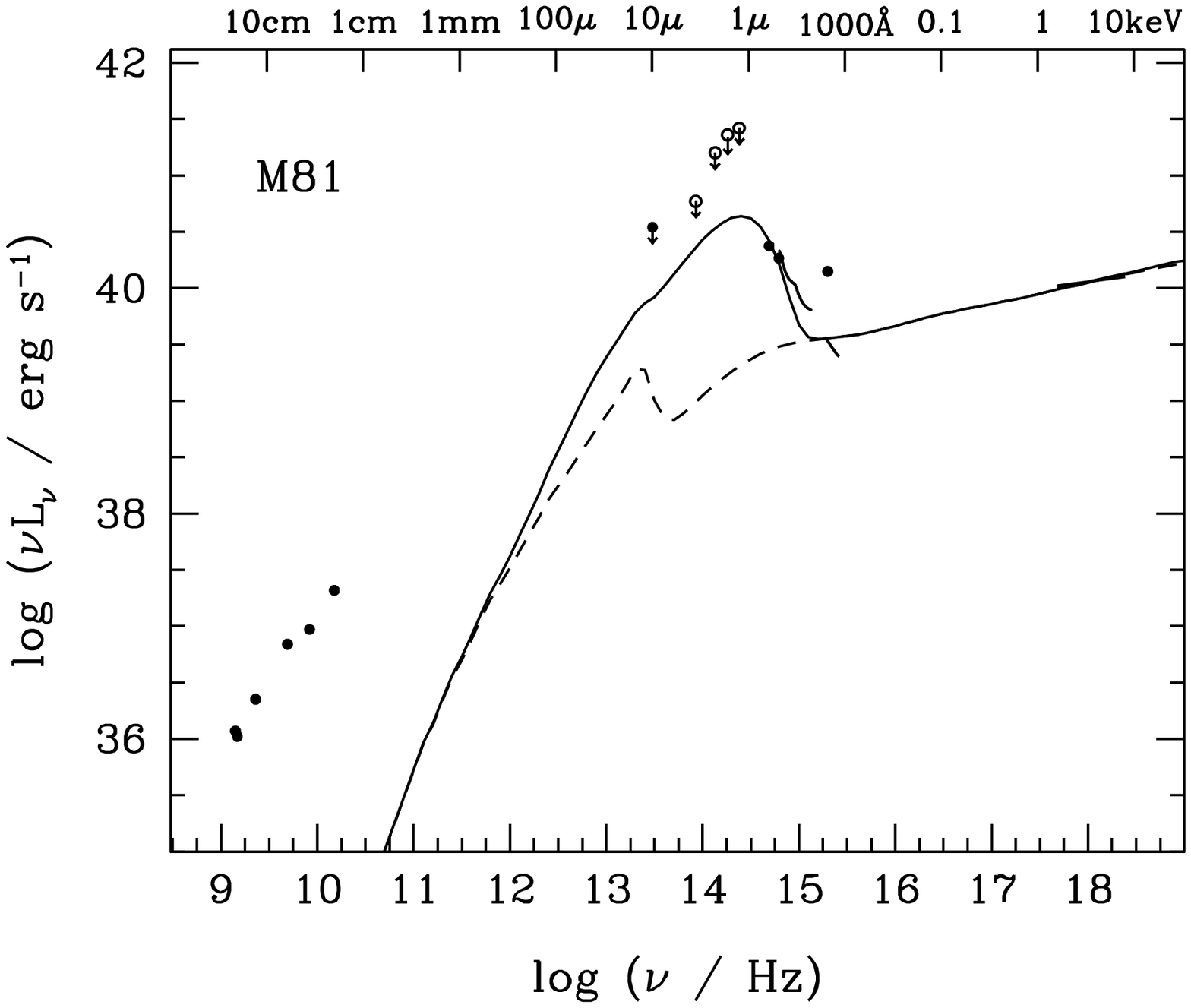}
\caption{{\bf Left:}  Multi-color blackbody thin accretion disk models for the 
  optical-UV emission from M81 (dotted lines from top to bottom: $\md
  = 10^{-2}, 10^{-3}, 3 \times 10^{-4}, \ \& \ 3 \times 10^{-5}$ with
  $\ri = 3$; solid line: $\md = 3 \times 10^{-3} \ \& \ \ri = 100$).
  {\bf Right:} A model for M81 in which a thin disk is truncated at
  $\ri \approx 100$, inside of which there is an ADAF.  The solid line
  shows the total ``disk + ADAF'' emission while the dashed line shows
  the ADAF contribution.  The truncated disk produces the optical/UV
  emission while the X-rays are produced in the ADAF.}

\end{figure}

\newpage

\begin{figure}
\plottwo{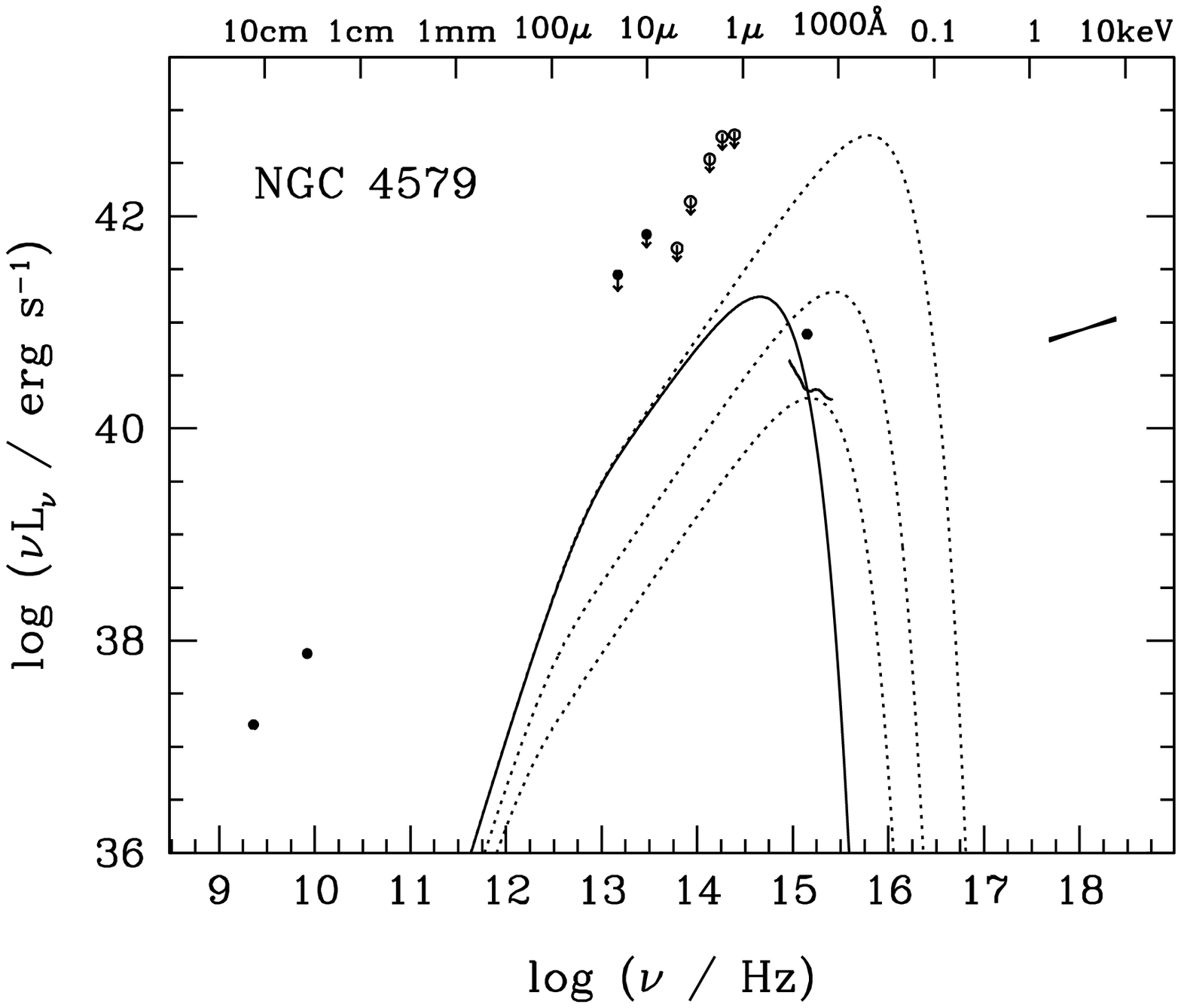}{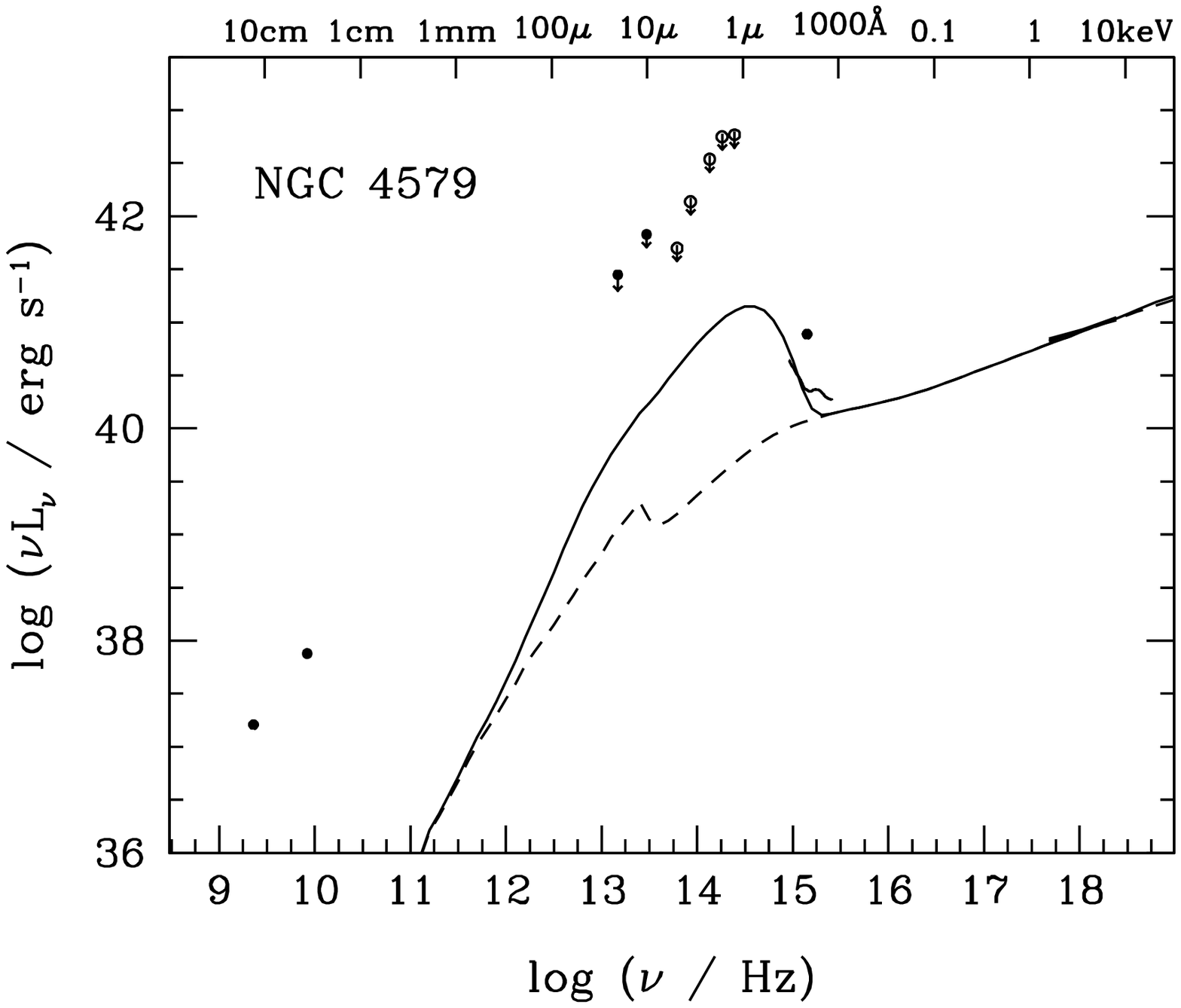}
\caption{{\bf Left:}  Multi-color blackbody thin accretion disk models for the 
  optical-UV emission from NGC 4579 (dotted lines from top to bottom:
  $\md = 3 \times 10^{-2}, 10^{-3}, \ \& \ 10^{-4}$ with $\ri = 3$;
  solid line: $\md = 0.03 \ \& \ \ri = 100$).  {\bf Right:} A model
  for NGC 4579 in which a thin disk is truncated at $\ri \approx 100$,
  inside of which there is an ADAF.  The solid line shows the total
  ``disk + ADAF'' emission while the dashed line shows the ADAF
  contribution.}
\end{figure}

%\end{multicols}

\begin{thebibliography}{991}
\bibitem{} Abramowicz, M., Chen, X., Kato, S., Lasota, J.P., \& Regev, O., 1995, ApJ, 438, L37 
%\bibitem{} Abramowicz, M., Czerny, B., Lasota, J. P., \& Szuszkiewicz, E., 1988, ApJ, 332, 646.
%\bibitem{} Allen, S., Di Matteo T., \& Fabian, A., 1999, MNRAS, submitted
%\bibitem{} Balbus, S. A., \& Hawley, J.F. 1998, Rev. Mod. Phys., 70, 1
\bibitem{} Barth, A. J. et al., 1996, AJ, 112, 1829
%\bibitem{} Begelman M. C., 1985, in {\it Astrophysics of Active Galaxies and Quasi-stellar Objects}, ed. Miller, J.S. (University Science Books: Mill Valley)
\bibitem{} Begelman, M. C., 1986, Nature, 322, 614
\bibitem{} Bietenholz, M. F. et al., 1996, ApJ, 457, 604
%\bibitem{} Begelman, M. C., 1986, Nature, 322, 614
%\bibitem{} Biretta, J., in {\it Astrophysical Jets}, ed. Burgarella, D., Livio, M., \& O'Dea, C. P., 1993  (Cambridge: Cambridge University Press)
%\bibitem{} Bisnovatyi-Kogan, G.S. \& Lovelace, R. V. E. 1997, ApJ, 486, L43 
%\bibitem{} Blackman, E., 1999, MNRAS, 302, 723 
%\bibitem{} Blandford, R. D. \& Begelman, M. C., 1999, MNRAS in press (astro-ph/9809083)
\bibitem{} Bondi, H., 1952, MNRAS, 112, 19
\bibitem{} Bower, G. A., Wilson, A.S., Heckman, T. M., \& Richstone, D.O., 1996, in {The Physics of LINERs in View of Recent Observations,} ed. M. Eracleous et al.. (San Francisco:  ASP), 163
%\bibitem{} Buote, D. \& Fabian, A., 1998, MNRAS, 296, 977
%\bibitem{} Di Matteo, T. \& Fabian, A., 1997, MNRAS, 286, L50
\bibitem{} Di Matteo, T., Fabian, A. C., Rees, M. J., Carilli, C. L., \& Ivison, R. J., 1999a, MNRAS, 305, 492 
\bibitem{} Di Matteo, T., Quataert, E., Allen, S., Narayan, R., \& Fabian, A. C., 1999b, MNRAS, submitted (astro-ph/9905053)
\bibitem{} Esin, A. A., McClintock, J. E., \& Narayan, R., 1997, ApJ, 489, 86 
%\bibitem{} Fabian, A. C. \& Canizares, C. R., 1988, Nature, 333, 829 
%\bibitem{} Fabian, A. C. \& Rees, M. J., 1995, MNRAS, 277, L55 
%\bibitem{} Fabbiano, G., \& Juda, J. Z., 1997, ApJ, 476, 666
%\bibitem{} Filippenko, A. V., 1996, in {\it The Physics of LINERs in View of Recent Observations}, eds. Eracleous, M., Koratkar, A., Leitherer, C., \& Ho, L. (ASP Conference Series Vol 103)
%\bibitem{} Ford, H. C. et al., 1994, ApJ, 435, L27
\bibitem{}  Frank, J., King, A., \& Raine, D., 1992, {\it Accretion Power in Astrophysics} (Cambridge Univ. Press)
\bibitem{} Gammie, C.F., Narayan, R., \& Blandford, R., 1999, ApJ, 516, 177
%\bibitem{} Goodman, J. \& Lee, H., 1989, ApJ, 337, 84
%\bibitem{} George, I. \& Fabian, A., 1991, MNRAS, 249, 352
%\bibitem{} Gruzinov, A. 1998, ApJ, 501, 787 
%\bibitem{} Gruzinov, A. 1999, ApJ, in press
\bibitem{} Haardt, F. \& Maraschi, L., 1991, ApJ, 380, L51
%\bibitem{} Harms, R.J. et al. 1994, ApJ, 435, L35
%\bibitem{} Hawley, J. F., Gammie, C. F., \& Balbus, S. A. 1996, ApJ, 464, 690
%\bibitem{} Herrnstein, J., Greenhill, L., Moran, J., Diamond, P., Inque, M., Nakai, N., and Miyoshi, M., 1998, ApJ, 497, L69
%\bibitem{} Ho, L. C., 1998, in {\it Observational Evidence for Black Holes in the Universe}, ed. S. K. Chakrabarti (Dordrecht:  Kluwer), 157
\bibitem{} Ho, L. C., 1999, ApJ, 516, 672  (H99)
\bibitem{} Ho, L. C., Filippenko, A. V., \& Sargent, W. L. W., 1996, ApJ, 462, 183
%\bibitem{} Ho, L. C., Filippenko, A. V., \& Sargent, W. L. W., 1997, ApJ, 487, 568
\bibitem{} Ho, L. C., Van Dyk, S. D., Pooley, G. G., Sramek, R. A., \& Weiler, K. W., 1999, AJ, in press (astro-ph/9905077)
\bibitem{} Honma, F. 1996, PASJ, 48, 77
\bibitem{} Ichimaru, S. 1977, ApJ, 214, 840 
\bibitem{} Ishisaki, Y. et al., 1996, PASJ, 48, 237
%\bibitem{} Kato, S., Fukue, J., Mineshige, S., 1998, {\em Black-Hole Accretion Disks} (Japan: Kyoto University Press) 
\bibitem{} Kato, S., Fukue, J., Mineshige, S., 1998, {\em Black-Hole Accretion Disks} (Japan: Kyoto University Press) 
\bibitem{} Koratkar, A. \& Blaes, O. 1999, PASP, 111, 1
%\bibitem{} Kormendy, J. et al., 1997, ApJ, 473, L91
%\bibitem{} Kormendy, J. \& Richstone, D. O., 1995, ARA\&A, 33, 581
%\bibitem{} Laor, A. \& Draine, B., 1993, ApJ, 402, 441
\bibitem{} Lasota, J.-P., Abramowicz, M. A., Chen, X., Krolik, J., Narayan, R.,\& Yi, I.,  1996, ApJ,  462, 142
%\bibitem{} Magorrian, J. et al., 1998, AJ, 115, 2285
%\bibitem{} Malkan, M. A. \& Sargent, W. L. W., 1982, ApJ, 254, 22
\bibitem{} Meyer, F. \& Meyer-Hofmeister, E., 1994, A\&A, 288, 175
%\bibitem{} Nandra, K. \& Pounds, K., 1994, MNRAS, 268, 405
%\bibitem{} Nandra et al., 1997, ApJ, 477, 602
%\bibitem{} Narayan, R., Mahadevan, R., Grindlay, J.E., Popham, R.G., \& Gammie, C., 1998a, ApJ, 492, 554
\bibitem{} Narayan, R., Mahadevan, R., \& Quataert, E., 1998, in {\em The Theory of Black Hole Accretion Discs}, eds. M.A. Abramowicz, G. Bjornsson, and J.E. Pringle (Cambridge:  Cambridge University Press) (astro-ph/9803131)
\bibitem{} Narayan, R., \& Yi, I., 1994, ApJ, 428, L13  
%\bibitem{} Narayan, R., \& Yi, I., 1995a, ApJ, 444, 231  
%\bibitem{} Narayan, R., \& Yi, I., 1995b, ApJ, 452, 710 
%\bibitem{} Nicholson, K. L. et al., 1998, MNRAS, 300, 893
\bibitem{} Poutanen, J., Krolik, J. H. \& Ryde, F.  1997, MNRAS, 292, L21 
\bibitem{} Quataert, E. \& Narayan, R., 1999, ApJ, 520, 298
%\bibitem{} Quataert, E., Narayan, R., \& Reid, M., 1999, ApJ Letters in press 
\bibitem{} Rees, M. J., Begelman, M. C., Blandford, R. D., \& Phinney, E. S., 1982, Nature, 295, 17 
%\bibitem{} Reynolds, C., Di Matteo, T., Fabian, A., Hwang, U., \& Canizares, C., 1996, MNRAS, 283, L111
\bibitem{} Serlemitsos, P., Ptak, A., \& Yaqoob, T., 1996, in {The Physics of LINERs in View of Recent Observations}, eds. Eracleous, M., Koratkar, A., Leitherer, C., \& Ho, L. C. (San Francisco:  ASP), 103
\bibitem{} Shakura, N. I., \& Sunyaev, R. A., 1973, A\&A, 24, 337 
\bibitem{} Shimuar, T. \& Takaraha, F., 1995, ApJ, 445, 780
%\bibitem{} Shields, G. A., 1978, Nature, 272, 706
\bibitem{} Terashima, Y. et al., 1998, ApJ, 503, 212
%\bibitem{} Tsvetanov, Z. et al., 1998, ApJ, 493, L83
%\bibitem{} Turner, T., Gorge, I., Nandra, K., \& Mushotzky, R., 1997, ApJS, 113, 23
\end{thebibliography}
\end{document}